\documentclass[onecolumn,secnumarabic,amssymb, amsmath, nofootinbib,tightenlines,
nobibnotes, aps, prl,epsfig]{revtex4}
\usepackage{graphicx}
\usepackage{dcolumn}
\usepackage{bm}
\begin{document}
\preprint{APS/123-QED}
\title{Entropy and DIS structure functions}

\author{G.R.Boroun}%
 \email{ boroun@razi.ac.ir }
\affiliation{ Department of Physics, Razi University, Kermanshah
67149, Iran}
\date{\today}
\begin{abstract}
Entanglement entropy in Deep Inelastic Scattering (DIS) from the
DIS structure functions has emerged as a novel tool for probing
observable quantities. The method proposed by Kharzeev-Levin to
determine entanglement entropy in DIS from parton distribution
functions (PDFs) improves on the momentum-space approach proposed
by Lappi et al.[Eur. Phys. J. C {\bf84}, 84 (2024)] and further
developed by Boroun and Ha [Phys. Rev. D {\bf109}, 094037 (2024)]
using Laplace transform techniques. The entropy of charged hadrons
is obtained from the parameterization of the proton structure
function and compared with H1 data, HSS, and HERA PDFs. Our
results for the entanglement entropy align very well with the H1
data across a wide range of $x$ and $Q^2$. Finally, the behavior
of the entanglement entropy is described at fixed $\sqrt{s}$ to
the minimum value of $x$ given by $Q^2/s$, which indicates that the
polarization of the exchanged photon for entropy determination is
transverse
 at this specific kinematic point. The effect of adding a simple higher
twist term to the description of entropy at low-$x$ and low-$Q^2$
values for comparison with HERA data is
investigated.\\
\end{abstract}
 \pacs{***}
\maketitle
\section{Introduction}

Deep Inelastic Scattering is the process in which a high-energy
lepton, such as an electron, scatters off a hadron, like a
proton, by exchanging a virtual photon. DIS revealed the partonic
structure of the proton, specifically quarks and gluons, and was crucial
in establishing Quantum Chromodynamics (QCD). Entropy in
DIS is viewed through an information-theoretic lens,  measuring  the
information loss or the number of configurations inside the
proton. The Shannon\footnote{In information theory, the entropy
of a random variable quantifies the average level of uncertainty
or information associated with the variable$^{,}$s potential
states or possible outcomes. The concept of information entropy
was introduced by C.Shannon based on a data communication system
composed of  three elements: a source of data, a communication
channel, and a receiver in the following literature:\\
1. Claude E.Shannon, Bell System Technical Journal {\bf27}, 379
(1948); {\bf27}, 623 (1948).} entropy for a probability distribution
($p_{1}$,...., $p_{n}$) is given by
\begin{eqnarray}\label{Shanon_eq}
S=-\sum_{n}p_{n}{\ln}p_{n},
\end{eqnarray}
where the eigenvalues $p_{n}$ are defined as the probability of a
state with $n$ partons/dipoles. The entanglement in the
electron-proton DIS process provides new insights into the
gluon-dominated structure and properties of high-energy hadrons.
The proton is a pure state of the QCD Hamiltonian, and a
measurement of its partonic content at some resolution scale
introduces a bipartition between resolved (A region) and
unresolved (B region) parts, which can be interpreted as the
entanglement entropy resulting from the entanglement between
regions $A$ and $B$  \cite{Kharzeev, Levin, Ryb, Caputa}. This
entanglement can be quantified by the von Neumann entropy as
\begin{eqnarray}\label{von Neumann_eq}
S=-\sum_{n}p_{n}{\ln}p_{n}=-\mathrm{Tr}(\rho_{A}\mathrm{\ln}\rho_{A}).
\end{eqnarray}
Indeed, the virtual photon with momentum transfer $q^2=-Q^2$ in
DIS only probes a portion of the proton's wave function, denoted as
$A$. The summation of the unobserved part of the wave function
localized in the region $B$, which is complementary to $A$, causes the
inclusive DIS measurements. Therefore, the reduced density matrix
is defined as $\widehat{\rho}_{A}=\mathrm{tr}_{B}\widehat{\rho}$,
where $\widehat{\rho}=|\psi{\rangle}{\langle}\psi|$ is the density
matrix of a hadron in its rest frame.\\
For a pure quantum mechanical state $|\psi{\rangle}$, the von
Neumann entropy  is  zero as
$S=-\mathrm{tr}[\widehat{\rho}{\ln}\widehat{\rho}]=0$ . The
composite system in $A{\cap}B$ (the entire proton) in the Hilbert
space $\mathcal{H}_{A}{\otimes}\mathcal{H}_{B}$ is described by
the pure wave function $|\Psi_{AB}{\rangle}$. Due to the Schmidt
decomposition theorem \cite{Schmidt}, it can be expanded as a
single sum
\begin{eqnarray}\label{Psi_eq}
|\Psi_{AB}{\rangle}=\sum_{n}\alpha_{n}|\Psi_{n}^{A}{\rangle}|\Psi_{n}^{B}{\rangle}.
\end{eqnarray}
The states $|\Psi_{n}^{A}{\rangle}$ and $|\Psi_{n}^{B}{\rangle}$
localized in the domains $A$ and $B$ are orthonormal and are given
by the Fock states with different numbers $n$ of partons in the
parton model. Now, the density matrix in region $A$ is defined by
the following form
\begin{eqnarray}\label{RhoA_eq}
\rho_{A}=\mathrm{tr}_{B}\rho_{AB}=\sum_{n}\alpha_{n}^{2}\int
|\Psi_{n}^{A}{\rangle}|\Psi_{n}^{B}{\rangle}{\langle}\Psi_{n}^{B}|{\langle}\Psi_{n}^{A}|dz
=\sum_{n}\alpha_{n}^{2}
|\Psi_{n}^{A}{\rangle}{\langle}\Psi_{n}^{A}|,
\end{eqnarray}
where the coordinate $z{\in}B$ and $p_{n}=\alpha_{n}^{2}$
\cite{Kharzeev}.\\
The proton at small $x$ (large rapidity $Y={\ln}\frac{1}{x}$
regime) is composed of an exponentially large number N of
micro-states that occur with equal and small probabilities of
${1}/{N}$ \cite{Levin}. The hadrons produced in the current
fragmentation region of DIS originate from the hadronization of
the struck quark, which is a constituent of the color-singlet
dipole, and the multiplicity of color-singlet dipoles. The
multiplicity distribution in QCD cascade follows the following
formula \cite{Tu}
\begin{eqnarray}\label{Pn eq}
p_{n}(N)=\frac{1}{N}\bigg(1-\frac{1}{N}\bigg)^{n-1},
\end{eqnarray}
where $N$ is the average multiplicity of color-singlet dipoles.
Therefore, the von Neumann entropy is defined by the following
formula
\begin{eqnarray}\label{SN eq}
S(x,Q^2)=-\sum_{n}p_{n}{\ln}p_{n}|_{N{\rightarrow}\infty}{\simeq}\ln{N}(x,Q^2),
\end{eqnarray}
where $N$ is the number of partons with longitudinal momentum
fraction $x$ \cite{Brodsky} through the equation
\begin{eqnarray}\label{Parton eq}
{N}(x,Q^2)=xg(x,Q^2)+x\Sigma(x,Q^2),
\end{eqnarray}
here $xg$ and $x\Sigma$ represent the gluon and singlet
distribution functions \cite{Kutak1}. Some pictures suggest that
partonic entropy coincides with the entropy of final state hadrons
in DIS \cite{Khoze, Mueller}. Since the partons cannot be isolated
as asymptotic states, the number of partons is not a directly
observable quantity.  Instead, it depends on the virtuality scale
of the process $Q^2$. Therefore, the quantum DIS entropy is not
directly observable \cite{Brodsky}.\\
Since the parton distribution functions themselves are not
physical observables, they have larger errors than the
experimental data from which they have been determined. The PDFs
resulting from fits to data at different orders in the running
coupling or different factorization schemes can appear very
different and are difficult to compare to each other in a
meaningful way \cite{Forte1, Forte2, Collins}. These different
functional forms can, in fact, lead to similar values for physical
observables. Since PDFs are not physical observables,  we
will define the entanglement entropy in DIS processes in terms of
observable quantities, such as DIS structure functions. The
longitudinal structure function in a momentum-space approach
\cite{Lappi, BH1, BH2} is defined as the proton structure
function, which is measurable in deeply inelastic scattering based
on the Block-Durand-Ha (BDH) parameterization \cite{Martin1}. The
momentum-space approach has two advantages: there is no need to
define a factorization scheme,and the approach in terms of
physical structure functions has the advantage of being more
transparent in the parametrization of the initial conditions of
the evolution. This parameterization applies to large and small
$Q^2$ and small $x$. The BDH  fit is crucial for accurately
describing the behavior of the proton structure function in
regions where data is sparse and aligns well with theoretical
predictions, such as the Froissart bound, which describes the
asymptotic behavior of hadron-hadron cross sections.\\
We examine entropy  due to the HERA center-of-mass (COM) energies,
then we will extend the ratio to the limit
$x_{\mathrm{min}}=Q^2/s$ which corresponds to high inelasticity
$y=1$, for comparison with HERA data. By including an additional
higher-twist term in the description of entropy $S(Q^2/s,Q^2)$,
the results will improve at low-$x$ and low-$Q^2$ values in
comparison with the HERA data.
 The main goal of the paper is entropy $S(Q^2/s,Q^2)$  with the COM energies
at the LHeC \cite{LHeC} and EIC \cite{EIC1, EIC2} colliders. This
ratio will define a limit bound for the
data in these colliders.\\

\section{Method}

The long-range physics in hard processes is related to the
nonperturbative structure of QCD bound states and can be defined
in terms of PDFs, which can, at least at leading order, be given
an intuitive physical interpretation as parton densities. The
short-range parts can be systematically calculated by the
perturbative techniques of QCD and electroweak theory. These
contributions are factorized into the PDFs, which then become
dependent on a factorization scale $\mu$, defining the
separation between  short and long-distance physics. The result
of a perturbative calculation at a fixed perturbative order
depends on the factorization scheme, typically chosen to be
the $\overline{\mathrm{MS}}$ scheme. \\
The PDFs, as non-observable quantities, have significant
parametrization freedom, leading to additional uncertainties
as different functional forms can yield similar values
for physical observables. Since PDFs are not physical observables,
it is unclear to what extent one should  favor smooth functions for PDFs, making the comparison between
calculations and measurements  less direct. In Ref.\cite{Lappi},
the authors define an alternative to the conventional
PDF-based approach in terms of observable quantities, which has
the advantage of reducing theoretical uncertainty.\\
The DIS structure functions, $F_{i}(x,Q^2)$ where $i=2$ or $L$,
and PDFs, $f_{j}(x,\mu^2)$ where $j=s$ or $g$, both
dependent on a factorization scale $\mu$, are related to each
other through scheme-dependent Wilson coefficient functions,
$C_{i,j}$, in the collinear factorization framework:
\begin{eqnarray}\label{FCf eq}
F_{i}(x,Q^2)=\sum_{j}C_{ij}(Q^2,\mu^2){\otimes}f_{j}(\mu^2).
\end{eqnarray}
The inverse of relation Eq.(\ref{FCf eq}) that expresses the PDFs in
terms of the structure functions is defined by the following form
\begin{eqnarray}\label{IFCf eq}
f_{i}(x,\mu^2)=\sum_{j}C^{-1}_{ij}(Q^2,\mu^2){\otimes}F_{j}(Q^2),
\end{eqnarray}
where $C^{-1}_{ij}$ denotes the perturbative inverse of $C_{ij}$.
The evolution equations for the dependence of physical observables
on the physical scale $\mu^2=Q^2$, without reference to PDFs, are
\begin{eqnarray}\label{DFCf eq}
\frac{dF_{i}(x,Q^2)}{d{\ln}Q^2}=\sum_{k}P_{ik}(Q^2){\otimes}F_{k}(Q^2),
\end{eqnarray}
where
\begin{eqnarray}\label{PCf eq}
P_{ik}=\sum_{j}\frac{d}{d{\ln}Q^2}C_{ij}(Q^2,\mu^2){\otimes}C^{-1}_{kj}(Q^2,\mu^2).
\end{eqnarray}
The DIS structure functions $F_{2}$ and $F_{L}$ for deeply
inelastic scattering into the PDFs are defined by the following
forms \cite{Lappi}
\begin{eqnarray} \label{Lappi1_eq}
F_{2}(x,Q^2)&=&<e^2>\bigg{\{}
C^{(0)}_{2,s}+\frac{\alpha_{s}(\mu^{2})}{2\pi}\bigg{[}C^{(1)}_{2,s}
-\ln{\bigg{(}}\frac{\mu^{2}}{Q^2}{\bigg{)}}C^{(0)}_{2,s}{\otimes}P_{qq}
\bigg{]}\bigg{\}}{\otimes}x\Sigma(x,\mu^{2})\nonumber\\
&&+2\sum_{i=1}^{n_{f}}e_{i}^{2}\frac{\alpha_{s}(\mu^{2})}{2\pi}\bigg{[}C^{(1)}_{2,g}
-\ln{\bigg{(}}\frac{\mu^{2}}{Q^2}{\bigg{)}}C^{(0)}_{2,g}{\otimes}P_{qg}
\bigg{]}{\otimes}xg(x,\mu^{2}),
\end{eqnarray}
and
\begin{eqnarray}\label{Lappi2_eq}
F_{L}(x,Q^2)&=&<e^2>\frac{\alpha_{s}(\mu^{2})}{2\pi}\bigg{\{}
C^{(1)}_{L,s}+\frac{\alpha_{s}(\mu^{2})}{2\pi}\bigg{[}C^{(2)}_{L,s}
-\ln{\bigg{(}}\frac{\mu^{2}}{Q^2}{\bigg{)}}C^{(1)}_{L,s}{\otimes}P_{qq}
-2n_{f}\ln{\bigg{(}}\frac{\mu^{2}}{Q^2}{\bigg{)}}C^{(1)}_{L,g}{\otimes}P_{gq}\bigg{]}\bigg{\}}{\otimes}x\Sigma(x,\mu^{2})\nonumber\\
&&+2\sum_{i=1}^{n_{f}}e_{i}^{2}\frac{\alpha_{s}(\mu^{2})}{2\pi}\bigg{\{}C^{(1)}_{L,g}+
\frac{\alpha_{s}(\mu^{2})}{2\pi}\bigg{[}C^{(2)}_{L,g}-\ln{\bigg{(}}\frac{\mu^{2}}{Q^2}{\bigg{)}}C^{(1)}_{L,s}{\otimes}P_{qg}
-\ln{\bigg{(}}\frac{\mu^{2}}{Q^2}{\bigg{)}}C^{(1)}_{L,g}{\otimes}P_{gg}
\bigg{]}\bigg{\}}{\otimes}xg(x,\mu^{2})\nonumber\\
&&+<e^2>\bigg{(}\frac{\alpha_{s}(\mu^{2})}{2\pi}\bigg{)}^2\bigg{[}b_{0}\ln{\bigg{(}}\frac{\mu^{2}}{Q^2}{\bigg{)}}
\bigg{]}\bigg{[}C^{(1)}_{L,s}{\otimes}x\Sigma(x,\mu^{2})+
2n_{f}C^{(1)}_{L,g}{\otimes}xg(x,\mu^{2}) \bigg{]}.
\end{eqnarray}
Here the average quark charge is $<e^2>=\sum{e^{2}_{q}}/n_{f}$
where $n_{f}$ is the number of massless flavors and $P_{ij}$'s are
splitting functions. The result of inverting of the Eqs.
(\ref{Lappi1_eq}) and (\ref{Lappi2_eq}) into the singlet and gluon
PDFs in terms of the DIS structure functions is defined by the
following forms
\begin{eqnarray} \label{Lappi3_eq}
x\Sigma(x,\mu^{2})&=&\frac{1}{<e^2>}F_{2}(x,Q^2),
\end{eqnarray}
and
\begin{eqnarray} \label{Lappi4_eq}
g(x,\mu^2)&=&\int_{x}^{1}\frac{dy}{y}\delta(1-\frac{x}{y})\bigg{\{}\eta\bigg{[}\frac{{\partial}F_{2}(y,Q^2)}{{\partial}y}-3\frac{F_{2}(y,
Q^2)}{y}\bigg{]} \nonumber \\
&&+\zeta\frac{2\pi}{\alpha_{s}(Q^2)}\bigg{[}{y}\frac{{\partial}^2F_{L}(y,
Q^2)}{{\partial}^2y}-4\frac{{\partial}F_{L}(y,
Q^2)}{{\partial}y}+\frac{6}{y}F_{L}(y, Q^2)\bigg{]} \bigg{\}}.
\end{eqnarray}
where $\eta={C_{F}}/({4T_{R}n_{f}<e^2>})$ and
$\zeta={1}/({8T_{R}n_{f}<e^2>})$ with the color factors
$T_{R}=1/2$ and $C_{F}=4/3$ associated with the color group SU(3).
By setting the renormalization scale equal to the momentum
transfer, $\mu^2=Q^2$, the gluon distribution in terms of the DIS
structure functions can be expressed using the delta function
property, as \cite{BH2}
\begin{eqnarray} \label{BH1_eq}
xg(x,Q^2)=\eta\bigg{[}x\frac{{\partial}F_{2}(x,Q^2)}{{\partial}x}-3{F_{2}(x,
Q^2)}\bigg{]}
+\zeta\frac{2\pi}{\alpha_{s}(Q^2)}\bigg{[}{x^2}\frac{{\partial}^2F_{L}(x,
Q^2)}{{\partial}^2x}-4x\frac{{\partial}F_{L}(x,
Q^2)}{{\partial}x}+{6}F_{L}(x, Q^2)\bigg{]}.
\end{eqnarray}
Therefore, the DIS entropy, which is not directly observable in
the PDFs, can be defined in the DIS structure functions, which are
observable. We find that
\begin{eqnarray} \label{BH2_eq}
S(x,Q^2)&=& {\ln}[xg(x,Q^2)(\mathrm{Eq}.(\ref{BH1_eq}))+x\Sigma(x,Q^2)(\mathrm{Eq}.(\ref{Lappi3_eq}))]\nonumber\\
&&=\ln{\bigg\{}\bigg{[}{\eta}x\frac{{\partial}F_{2}(x,Q^2)}{{\partial}x}-\bigg(3\eta-\frac{1}{<e^2>}\bigg){F_{2}(x,
Q^2)}\bigg{]}\nonumber\\
&&+\zeta\frac{2\pi}{\alpha_{s}(Q^2)}\bigg{[}{x^2}\frac{{\partial}^2F_{L}(x,
Q^2)}{{\partial}^2x}-4x\frac{{\partial}F_{L}(x,
Q^2)}{{\partial}x}+{6}F_{L}(x, Q^2)\bigg{]}\bigg\}.
\end{eqnarray}
On the other hand, the authors in Ref.\cite{BH1} developed a
method to obtain the longitudinal structure function, $F_{L}$, in
the proton structure function and its derivative using a
Laplace-transform method detailed in \cite{Block1, Block2, Block3,
Block4, Block5}. Therefore, the momentum space evolution of the
longitudinal structure function $F_{L}(x,Q^2)$  from the proton
structure function and its derivative with respect to $\ln{Q^2}$
can then be written as
\begin{eqnarray} \label{BH3 eq}
F_{L}(x,Q^2)&=&4\int_{x}^{1}\frac{d{F}_{2}(z,Q^2)}{d{\ln}Q^2}(\frac{x}{z})^{3/2}\bigg{[}\cos{\bigg{(}}\frac{\sqrt{7}}{2}{\ln}\frac{z}{x}{\bigg{)}}-\frac{\sqrt{7}}{7}
\sin{\bigg{(}}\frac{\sqrt{7}}{2}{\ln}\frac{z}{x}{\bigg{)}}\bigg{]}\frac{dz}{z}-4C_{F}\frac{\alpha_{s}(Q^2)}{2\pi}\nonumber\\
&&{\times}\int_{x}^{1}F_{2}(z,Q^2)(\frac{x}{z})^{3/2}\bigg{[}(1.6817+2\psi(1))\cos{\bigg{(}}\frac{\sqrt{7}}{2}{\ln}\frac{z}{x}{\bigg{)}}
+(2.9542-2\frac{\sqrt{7}}{7}\psi(1))
\sin{\bigg{(}}\frac{\sqrt{7}}{2}{\ln}\frac{z}{x}{\bigg{)}}\bigg{]}\frac{dz}{z}\nonumber\\
&&+8C_{F}\frac{\alpha_{s}(Q^2)}{2\pi}
\sum_{k=1}^{\infty}\frac{k}{(k+1)^2-3(k+1)+4}\int_{x}^{1}F_{2}(z,Q^2)(\frac{x}{z})^{k+1}
\frac{dz}{z}.
\end{eqnarray}
The authors in Refs. \cite{Kotikov, B1, B2, B3, B4} derived
elegant formulas for the longitudinal structure function
$F_{L}(x,Q^2)$. This effect is of order $\alpha_{s}(Q^2)$, and is
represented as a convolution integral over $F_{2}(x,Q^2)$ and the
gluon density $xg(x,Q^2)$. The authors in Ref.\cite{Kotikov}
revived the parametrization of the longitudinal structure function
using the Mellin transform method.  They based their
parametrization on the proton structure function $F_{2}(x,Q^{2})$
as suggested by the authors in Ref.\cite{Martin1},  who performed
a fit to HERA data on deep-inelastic lepton-nucleon scattering
(DIS) at small vales of $x$.
\begin{eqnarray}\label{FL eq}
F_{L}(x,Q^{2})&=&\tau(a_{s})\{ \vartheta(a_{s})(1-
x)^{n}\sum_{\varepsilon=0}^{2}C_{\varepsilon}(Q^{2})L^{\varepsilon}-\chi(a_{s}^{2})F_{2}(x,Q^{2})\},
\end{eqnarray}
with
\begin{eqnarray}\label{FLC eq}
\tau(a_{s})& =&[1+\frac{1}{3}a_{s}(Q^{2})L_{C}
(\widehat{\delta}^{(1)}_{sg}-\widehat{R}^{(1)}_{L,g})]^{-1}\nonumber\\
\vartheta(a_{s})&=&[1-a_{s}(Q^{2})
(\overline{\delta}^{(1)}_{sg}-\overline{R}^{(1)}_{L,g})]\nonumber\\
\chi(a_{s}^{2})&=&
a^{2}_{s}(Q^{2})[\frac{1}{3}\widehat{B}^{(1)}_{L,s}L_{A}+\overline{B}^{(1)}_{L,s}].
\end{eqnarray}
The parametrization of the structure function $F_{ 2}(x,Q^{2})$ in
\cite{Martin1} is obtained from a combined fit to HERA data across
a wide range of kinematic variables $x$ and $Q^2$. The equation
for $F_{ 2}(x,Q^{2})$ is given by
\begin{eqnarray}\label{F2 eq}
F_{ 2}(x,Q^{2})& =&
D(Q^{2})(1-x)^{n}\sum_{\varepsilon=0}^{2}A_{\varepsilon}(Q^{2})L^{\varepsilon}.
\end{eqnarray}
Here, the $L^{,}$s represent logarithmic terms. The coefficient
functions
are detailed in Appendix A, and the effective parameters are defined in Table I.\\

In conclusion, the DIS entropy is directly defined in the proton
structure function as an observable. Indeed,
\begin{eqnarray} \label{SF2 eq}
S(x,Q^{2})= \ln[f(F_{2}(x,Q^2))].
\end{eqnarray}
The resulting partonic entropy agrees with the hadronic entropy if
we consider the hadronic multiplicity distribution as determined
from charged hadrons. The ratio of charged hadrons to total hadron
multiplicity assumes that the total number of produced hadrons is
approximately $3/2$ times the number of charged hadrons observed
in experiments. According to \cite{Kutak2}, the number of gluons
and seaquarks that yield charged hadrons is roughly $2/3$ of the
total parton number. Therefore, the DIS entropy for charged
hadrons is defined by the following form
\begin{eqnarray} \label{SF3 eq}
S(x,Q^{2})= \ln[f(F_{2}(x,Q^2))]+\ln(2/3).
\end{eqnarray}
For entropy  working in terms of observable quantities (i.e,
the DIS structure function), rather than parametrizing and fitting
unobservable PDFs, Eq. (\ref{SF3 eq}) provides an unambiguous way
to confront predictions of
entropy with experimental measurements \cite{Kutak3}.\\
In Ref.\cite{Taylor}, the author persisted statistical
tensions with the standard model in the low $Q^2$ HERA deep
inelastic scattering neutral current data characterized by a
turnover of the neutral current reduced cross section at low $x$
and low $Q^2$. One important experimental signature that sheds
light on this low $Q^2$ region is the expected $F_L$ to be small
because its dominant gluon component is strongly suppressed and
the polarization of the exchanged photon is transverse at this
kinematic point. Therefore, we expect that the reduced cross
section at low $x$ and low $Q^2$ region is proportional to the
proton structure function as $ \sigma_{r}{\simeq}F_{2}$ where
entropy is now related to the reduced cross section measurable in deeply inelastic scattering \cite{B5}. Now, it is
worth mentioning that the DIS entropy will be considered in the
limit where $x_{\mathrm{min}}=Q^2/s$, indicating that the
longitudinal polarization of the virtual photon at $y=1$ is zero.
We can conclude that the entropy is defined by the following form
at this limit,
\begin{eqnarray} \label{Entropyxmin eq}
S(x,Q^{2}){\rightarrow}S(Q^2/s,Q^{2}),
\end{eqnarray}
where $s$ is the center-of-mass energy. We propose a significant
experimental signature that illuminates the low $Q^2$ region for
determining the DIS entropy based on the DIS structure functions
of the hadronic multiplicity distribution. This determination is
carried out using charged hadrons at a fixed $\sqrt{s}$ to the
minimum value of $x$ given by $Q^2/s$.\\
Higher twist (HT) corrections in deeply inelastic scattering
within the saturation model are defined in the literature
\cite{HT1, HT2}. A twist analysis of the nucleon structure
functions $F_{T}$ and $F_{L}$ at small values of the Bjorken
variable $x$ reveals that for $F_{L}$, the higher twist
corrections are significant, while for $F_{2}=F_{T }+F_{L}$ there
is nearly complete cancellation. The HT corrections are
parameterized as a phenomenological unknown function, and the
values of the unknown parameters are determined through fits to
experimental data \cite{HT5, HT6, HT7}. It is common practice to
adjust the leading-twist structure function by incorporating a
term inversely proportional to $Q^2$ as a phenomenological power
correction to account for HT effects in the structure function:
\begin{eqnarray} \label{HT eq}
F^{HT}_{2}(x,Q^2)=F_{2}^{LT}(x,Q^2)\left(1+\frac{H_{2}(x)}{Q^2}
\right).
 \end{eqnarray}
Here,  $F_{2}^{LT}$ represents the leading-twist contribution to
$F_{2}$, while $F_{2}^{HT}$ denotes the structure function $F_{2}$
with higher-twist corrections included. In the studies by authors
in Refs.\cite{HT5, HT6, HT7}, the inclusion of an HT term in
$F_{2}$ at low-$x$ and low-$Q^2$ was investigated, revealing the
necessity for $H_{2}=0.12{\pm}0.07~\mathrm{GeV}^2$ for  data with
$y<1$. Therefore, the DIS entropy for charged hadrons with the HT
corrections is rewritten by the following form
\begin{eqnarray} \label{SF3HT eq}
S^{HT}(x,Q^{2})= \ln[f(F^{HT}_{2}(x,Q^2))]+\ln(2/3),
\end{eqnarray}
and at the limit $x_{\mathrm{min}}=Q^2/s$ it is,
\begin{eqnarray} \label{EntropyxminHT eq}
S^{HT}(x,Q^{2}){\rightarrow}S^{HT}(Q^2/s,Q^{2}).
\end{eqnarray}

\section{Results and Conclusion}

With the explicit form of the entanglement entropy [i.e.,
Eqs.(\ref{SF2 eq}), (\ref{SF3 eq}) and (\ref{Entropyxmin eq})], we
begin to extract the numerical results at small $x$ in a wide
range of $Q^2$ values, using the parametrization of $F_{2}(x,Q^2)$
given by Eq.(\ref{F2 eq}). In Fig.1, we have plotted the
entanglement entropy $S(x,Q^2)$ as a function of $x$
($1{\times}10^{-4}<x<5{\times}10^{-2}$) from Eq.(\ref{SF2 eq}) for
the DIS entropy and from Eq.(\ref{SF3 eq}) for the entropy of
charged hadrons.  The values of $Q^2$ used were $7.5, 15, 30$, and
$70~\mathrm{GeV}^2$, respectively. The figure also includes data
from the H1 Collaboration (H1 2021 \cite{H1A}) along with total
errors based on measurements of charged particle multiplicity
distributions in DIS at HERA, showing implications for the
entanglement entropy of partons. The H1 Collaboration reported
charged particle multiplicity distributions in positron-proton
(ep) DIS at a center-of-mass energy $\sqrt{s}=319~\mathrm{GeV}$.
The phase space for the measurement of multiplicity distributions
is defined in bins of the virtuality of the photon
$5<Q^2<100~\mathrm{GeV}^2$ and the inelasticity $y$ variable
$0.0375<y<0.6$. The entanglement entropy extracted from the
final-state hadron entropy Eq.(\ref{SF3 eq}), calculated from
charged particle multiplicity distributions, at various $Q^2$
values, is in good agreement with experimental data across a wide
range of $x$. The error bands in the entanglement entropy results
correspond to the uncertainty in the parametrization of
$F_{2}(x,Q^2)$ in \cite{Martin1}.\\
As shown in the figure, the results from Eq.(\ref{SF3 eq}) at
average $Q^2$ values are comparable to the H1 data in $Q^2$ bins
across a wide range of $x$. These results are compared to the the
total entropy of the dipole model:
\begin{eqnarray} \label{RegeeE eq}
S(x,Q^{2})=\ln\left[C\left(\frac{1}{x}\right)^{\Delta}\right]+\ln(2/3).
\end{eqnarray}
In this equation, the Regge behavior is directly proportional to
the average number of dipoles and the intercept $\Delta$ is
determined from a fit to the parton distribution function at low
$x$ in \cite{Kutak2}. The authors in Ref.\cite{Kutak3} used
leading order HERAPDFs \cite{HZ} and PDFs obtained from the HSS
unintegrated gluon distribution \cite{HSS} for the parton
distributions. The results of the fit for the $C$ parameter and
$\Delta$ intercept are presented in Table II, where the
uncertainties are attributed to the choice of the hard scale and
 a variation in the range of $x$. The results of these fits (HSS and HERAPDFs) are
 compared with the H1 data and our results in Fig.1. We observe
 that the behavior of the H1 data is comparable to our curve
 results
 due to the behavior of the proton structure function
 compared with the linear behavior of the fits.
 However, we do observe a considerable quadratic behavior in our results
 and H1 data  compared to the linear behavior
 in the case of HSS PDFs and HERAPDFs.\\
\begin{figure}
\includegraphics[width=0.8\textwidth]{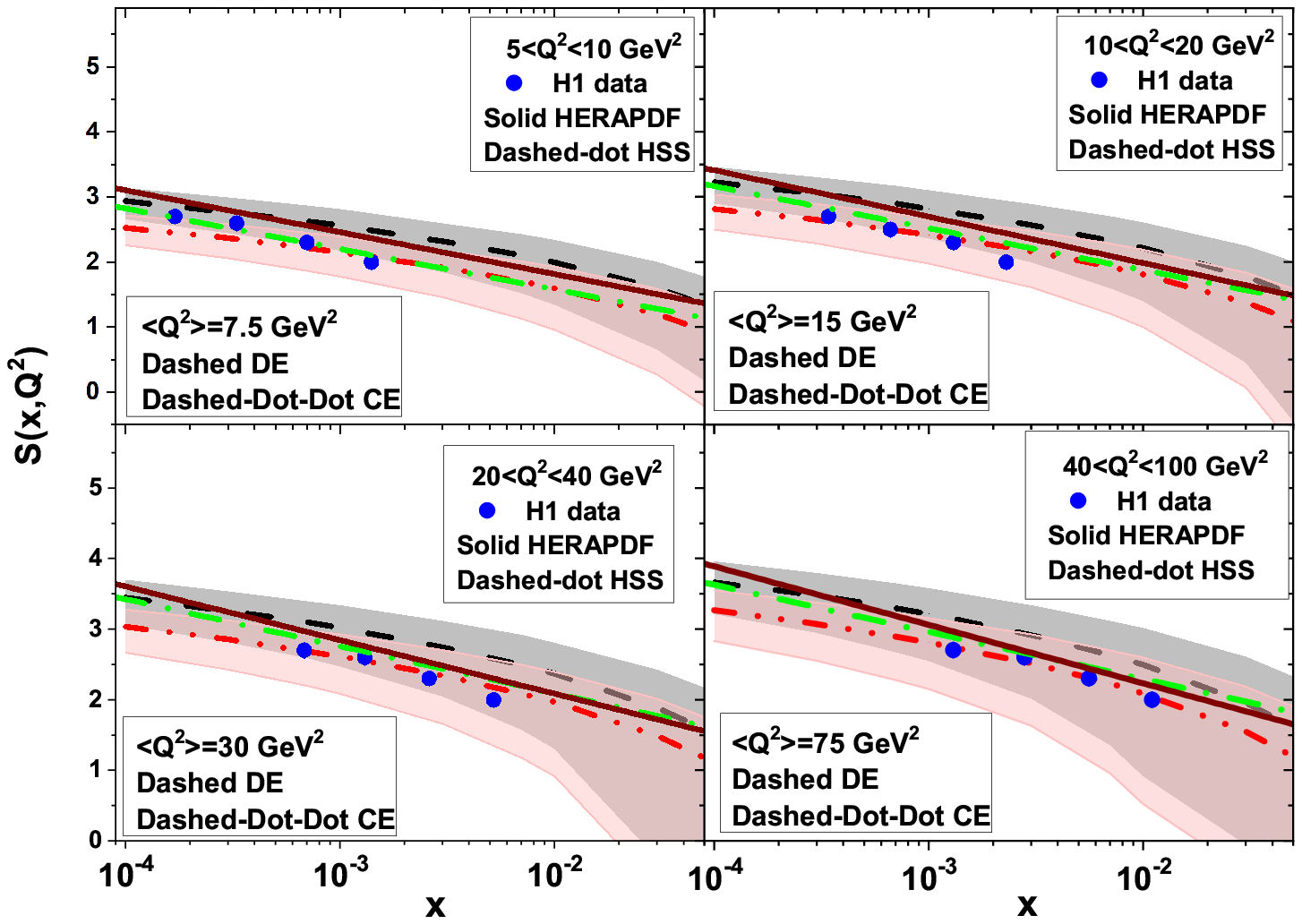}
\caption{The entanglement entropy results for the DIS entropy (DE,
black-dashed curve, Eq.(\ref{SF2 eq})) and the entropy of charged
hadrons (CE, red-dashed-dot-dot curve, Eq.(\ref{SF3 eq})) with
respect to the parametrization of $F_{2}(x,Q^2)$ are compared with
the H1 experimental data (blue-circles) \cite{H1A} as accompanied
with total errors and the results from HSS (green-dashed-dot
curve) and HERAPDF (brown-solid curve) as shown in \cite{Kutak3}.
The error bands are a result of  the parametrization of
$F_{2}(x,Q^{2})$. }\label{Fig1}
\end{figure}
In Fig.2, we show a comparison between the entanglement entropy
for the DIS entropy (Eq.(\ref{SF2 eq})) and the entropy of charged
hadrons (Eq.(\ref{SF3 eq})) with the H1 Collaboration data of the
Bjorken variable $x$ ($x=10^{-3}$) and in a wide range of $Q^2$.
Figure 2 clearly demonstrates that the entropy of charged hadrons
with respect to the proton structure function provides accurate
behaviors of the extracted H1 data. As seen in this figure, the
results are comparable to the H1 data  at all $Q^2$ values. We
observed that rescaling the entropy by a factor of $2/3$ improved
the results when comparing them to the measurements of charged
particle multiplicity distributions in DIS at HERA.\\
Indeed, the entanglement entropy based on the parametrization of
the structure function in deeply inelastic scattering has some
importance compared to  the Regge behavior of PDFs. In the
parametrization of $F_{2}$ that applies to large and small $Q^2$
and small $x$, there is an asymptotic behavior
($W^2{\rightarrow}\infty$ with $Q^2$ fixed), where
$F_{2}(W^2,Q^2){\propto}\ln^{2}(W^2/Q^2){\simeq}\ln^{2}(1/x)$,
with $W$ being the invariant mass.\\
\begin{figure}
\includegraphics[width=0.56\textwidth]{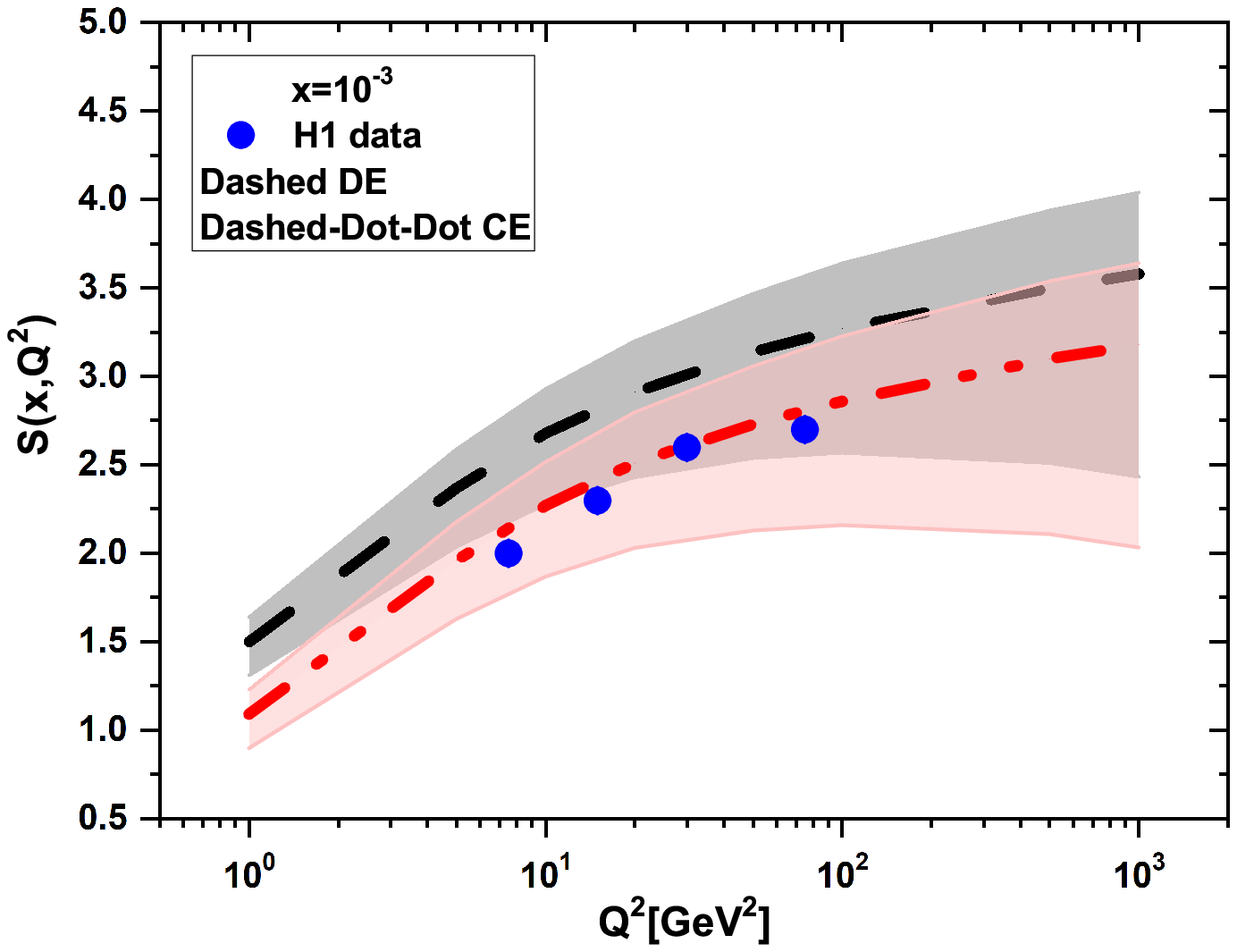}
\caption{The entanglement entropy results are compared at a fixed
value of the Bjorken variable $x$ ($x=10^{-3}$) and in a wide
range of $Q^2$ for the DIS entropy (DE, black-dashed curve,
Eq.(\ref{SF2 eq})) and the entropy of charged hadrons (CE,
red-dashed-dot-dot curve, Eq.(\ref{SF3 eq})) with respect to the
parametrization of $F_{2}(x,Q^2)$ in comparison to the H1
experimental data (blue-circles) \cite{H1A} along with total
errors. The error bands are a result of the parametrization of
$F_{2}(x,Q^{2})$.}\label{Fig2}
\end{figure}
In Fig.3, we consider HT corrections as multiplicative shifts to
the DIS structure function $F_{2}$, represented by the entropy
$S(\frac{Q^2}{s}, Q^2)$ in Eqs. (\ref{SF3HT eq}) and
(\ref{EntropyxminHT eq}), which  depend on
${F_{2}}(\frac{Q^2}{s}, Q^2)$. In this figure (i.e., Fig.3), the
entropy $S(\frac{Q^2}{s}, Q^2)$ is plotted against $Q^2$ values at
$x_{\mathrm{min}}=Q^2/s$ with and without HT corrections at
$\sqrt{s}=319~\mathrm{GeV}$  for $y=1$, accompanied by
uncertainties dependent on the model. We observe that the impact
of HT contribution on the ratio $\frac{S^{HT}}{S}(\frac{Q^2}{s},
Q^2)$
is noticeable at low $Q^2$ values.\\
\begin{figure}
\includegraphics[width=0.56\textwidth]{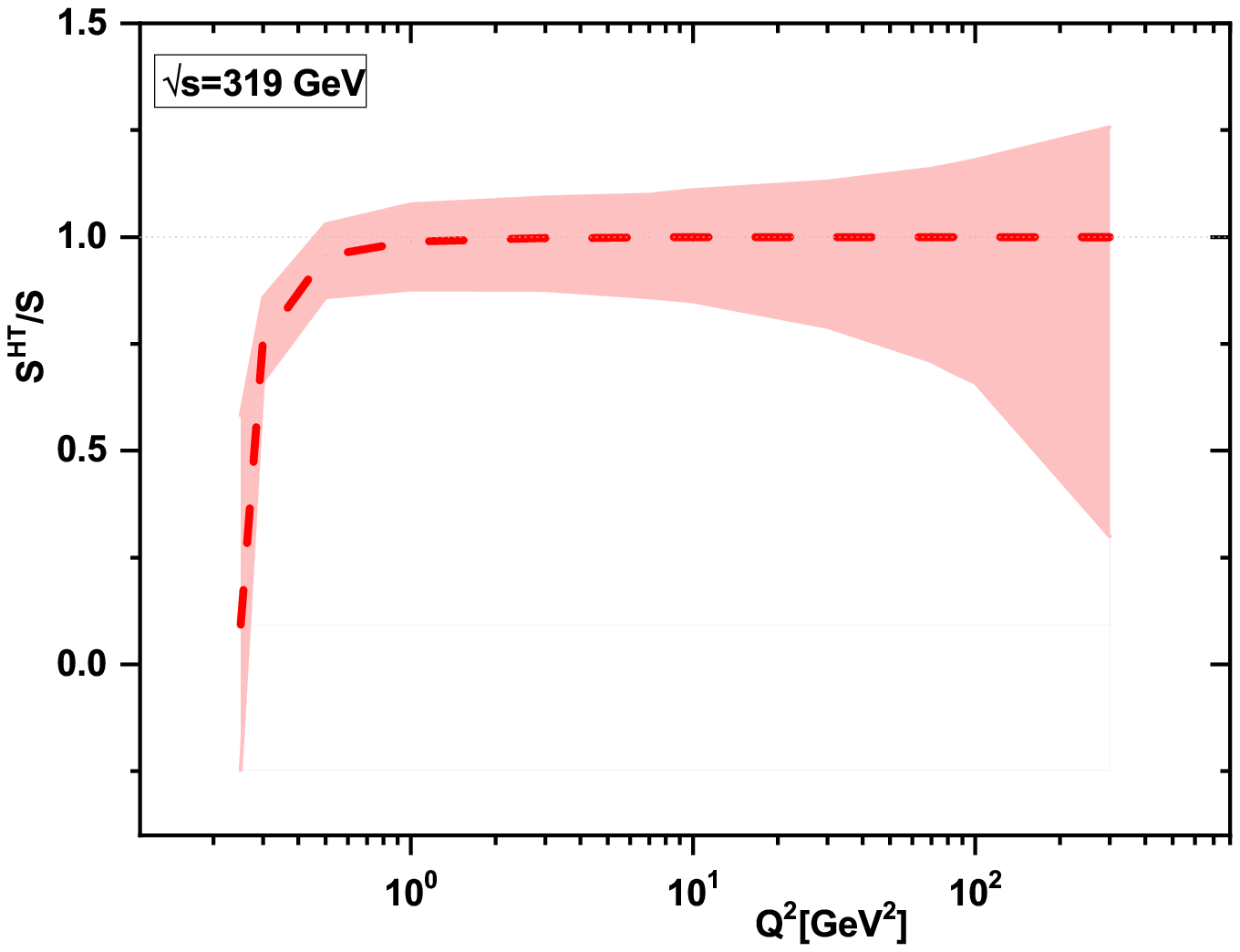}
\caption{The ratio of the HT-corrected entropy to the entropy
without HT corrections, $\frac{S^{HT}}{S}$, is shown as a function
of $Q^2$ at $x_{\mathrm{min}}=Q^2/s$ for
$\sqrt{s}=319~\mathrm{GeV}$ according to the HT corrections of
entropy of charged hadrons ( red-solid curve,
Eq.(\ref{EntropyxminHT eq})). The error bands result from the
parametrization of $F_{2}(x,Q^{2})$ and HT
uncertainties.}\label{Fig3}
\end{figure}
\begin{figure}
\includegraphics[width=0.56\textwidth]{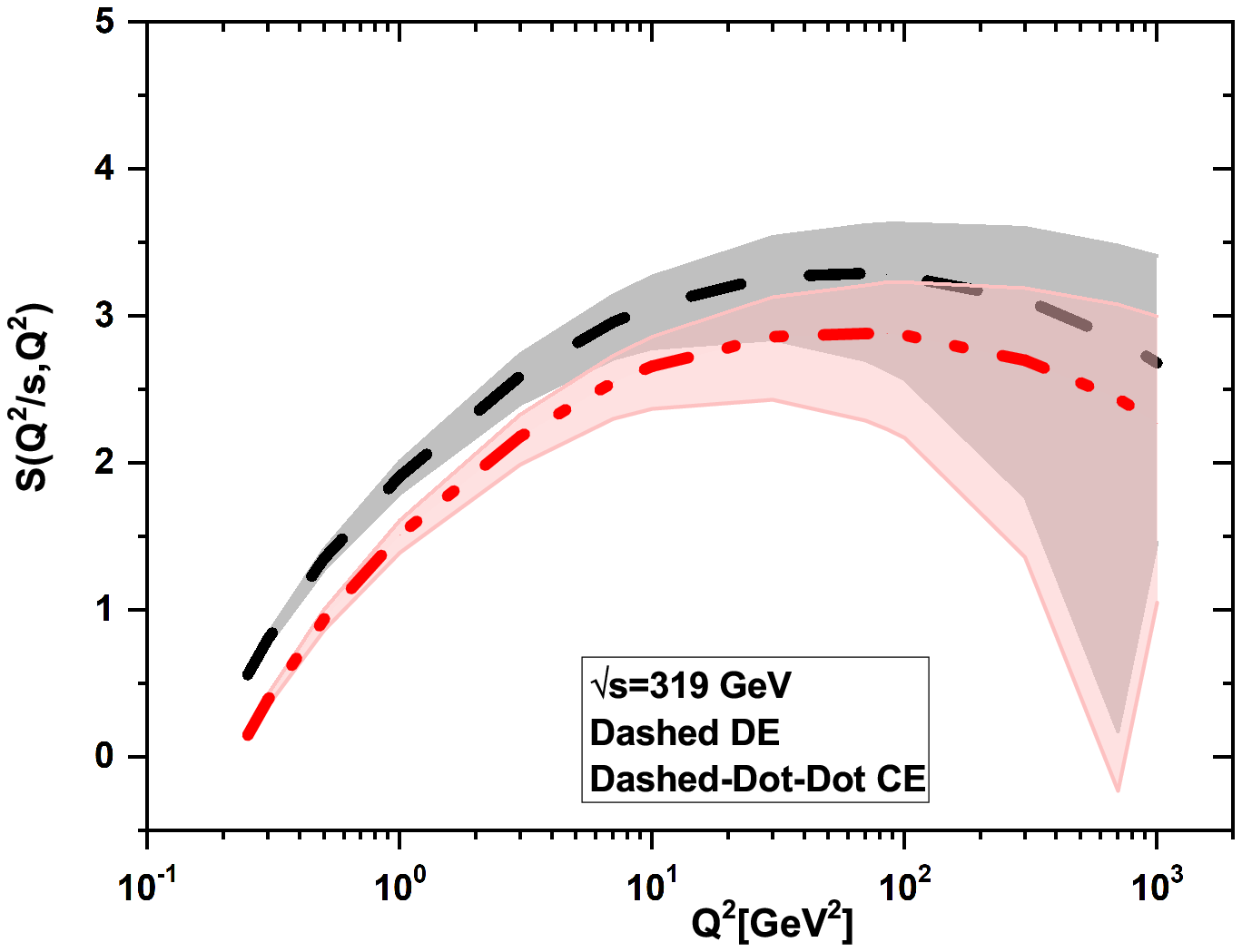}
\caption{The entanglement entropy extracted is shown as a function
of $Q^2$ at $x_{\mathrm{min}}=Q^2/s$ for
$\sqrt{s}=319~\mathrm{GeV}$ according to the DIS entropy (DE,
black-dashed curve, Eq.(\ref{SF2 eq})) and the entropy of charged
hadrons (CE, red-dashed-dot-dot curve, Eq.(\ref{SF3 eq})). The
error bands result from the parametrization of
$F_{2}(x,Q^{2})$.}\label{Fig4}
\end{figure}
In Fig.4, the entanglement entropies are obtained based on the
parametrization of the DIS structure functions at a fixed
$\sqrt{s}$ (i.e., $\sqrt{s}=319~\mathrm{GeV}$) at $y=1$, which
corresponds to $x_{\mathrm{min}}=Q^2/s$. We observe that at this
kinematic point, where the polarization of the virtual exchanged
photon is transverse, the entropy of charged hadrons (Eq.(\ref{SF3
eq})) is expected to decrease to $S(Q^2/s,Q^2){\rightarrow}0$,
otherwise it remains sizable for the DIS entropy (Eq.(\ref{SF2
eq})). We observe that at $x_{\mathrm{min}}=Q^2/s$ where the value
of $y$ becomes 1, the maximum entanglement entropy for the DIS
entropy and the entropy of charged hadrons reach  approximately
3.5 and 3, respectively, with corresponding uncertainties in the
interval of $20<Q^2{\lesssim}100~\mathrm{GeV}^2$. In this figure,
we observe that the significant feature of the entanglement
entropy versus $Q^2$ is the increase for $Q^2<20~\mathrm{GeV}^2$,
followed by a plateau from $Q^2{\sim}20$ to $100~\mathrm{GeV}^2$,
and then a decrease for larger values of $Q^2$. The QCD evolution
of the entanglement entropy \cite{Kutak4, Kutak5, B6} with
increasing $Q^2$, as well as the decline with increasing
$x_{\mathrm{min}}$, determines the overall shape.\\
The entanglement entropy as a function of $Q^2$ at
$x_{\mathrm{min}}$ is consistent with a second-order polynomial
fit in $\ln(Q^2)$. In order to determine the best values of the
fit terms, we use this function into $x_{\mathrm{min}}$ by the
following form at $\sqrt{s}=319~\mathrm{GeV}$ the entropy of
charged hadrons as
\begin{eqnarray} \label{Fit eq}
S(x_{\mathrm{min}},sx_{\mathrm{min}}){|}_{y=1}=-1.986-1.325~{\ln}(x_{\mathrm{min}})-0.0892~{\ln}^{2}(x_{\mathrm{min}}).
\end{eqnarray}
The results of the entropy of charged hadrons at
$x_{\mathrm{min}}=Q^2/s$ for $\sqrt{s}=319~\mathrm{GeV}$,
$S(x_{\mathrm{min}},sx_{\mathrm{min}})$, are shown in Fig.5. The
uncertainties of the entropy (dashed curves) have been included
due to errors in the parametrization of the proton structure
function coefficients. It has been observed that the maximum
entropy of charged hadrons at $y=1$ for
$\sqrt{s}=319~\mathrm{GeV}$ is approximately 3 for
$x{\simeq}10^{-3}$.\\
The derivative of $S(x_{\mathrm{min}},sx_{\mathrm{min}})$ changes
sign at $x{\simeq}4{\times}10^{-3}$ and is given by
\begin{eqnarray} \label{Delta eq}
\Delta(x_{\mathrm{min}},sx_{\mathrm{min}}){|}_{y=1}=\frac{d}{d{\ln}(\frac{1}{x_{min}})}
S(x_{\mathrm{min}},sx_{\mathrm{min}}){|}_{y=1}=
1.325+0.1784~{\ln}(x_{\mathrm{min}}).
\end{eqnarray}
We observe in Fig.6 that the values of
$\Delta(x_{\mathrm{min}},sx_{\mathrm{min}})$ depend on $x$ and
have both positive and negative values at the maximum and minimum
values of $x$. The negative values of $\Delta$ predict the zero
value of entanglement entropy as $Q^2{\rightarrow}0$. The HSS
\cite{HSS} and HERAPDF \cite{HZ} predicted values are constant, as
shown in Fig.6. However, the value of $\Delta_{\mathrm{HSS}}$ is
the lowest intercept value in Eq.(\ref{RegeeE eq}) within the bin
values of $Q^{2}= [5-10~\mathrm{GeV}^2]$, while the largest value
of $\Delta_{\mathrm{HERAPDF}}$ is within the bin values of $Q^{2}=
[40-100~\mathrm{GeV}^2]$. We observe that the evolution of
entanglement entropy is determined by an effective intercept in
Fig.6.\\
\begin{figure}
\includegraphics[width=0.56\textwidth]{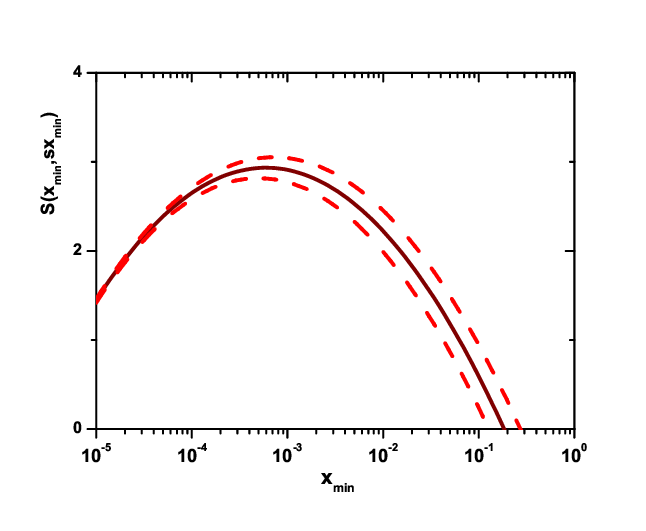}
\caption{The entanglement entropy extracted is shown as a function
of $x_{\mathrm{min}}$ for $\sqrt{s}=319~\mathrm{GeV}$ based on the
entropy of charged hadrons (red-solid curve). The error bands
result from the parametrization of $F_{2}(x,Q^{2})$ (red-dashed
curves).}\label{Fig5}
\end{figure}

\begin{figure}
\includegraphics[width=0.56\textwidth]{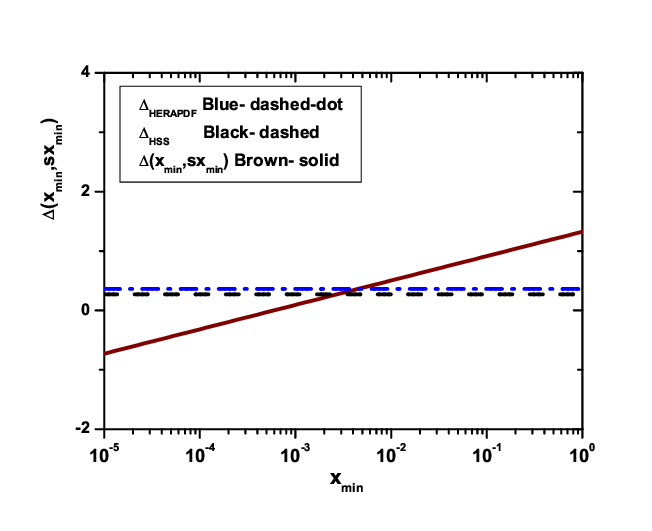}
\caption{The figure displays the effective intercept behavior of
the entanglement entropy extracted as a function of
$x_{\mathrm{min}}$ for $\sqrt{s}=319~\mathrm{GeV}$. It is based on
the entropy of charged hadrons (brown-solid curve) and is compared
with the HSS \cite{HSS} (black- dashed curve) and HERAPDF
\cite{HZ} (blue-dashed-dot curve) intercepts within the bin values
of $Q^{2}= [5-10~\mathrm{GeV}^2]$ and $Q^{2}=
[40-100~\mathrm{GeV}^2]$, respectively.}\label{Fig6}
\end{figure}
Future measurements at the EIC and LHeC will be examined at
upcoming colliders, serving as an interesting tool to investigate
entanglement entropy at low $x$ and moderate $Q^2$. Our
predictions for entanglement entropy $S(\frac{Q^2}{s}, Q^2)$ in
the kinematic range of the EIC and LHeC colliders are shown in
Fig.7. In this figure, the behavior of the entanglement entropy
$S(\frac{Q^2}{s}, Q^2)$ is considered with COM energies
$\sqrt{s}=89~\mathrm{GeV}$ and $1.3~\mathrm{TeV}$ for the EIC and
LHeC colliders respectively. The uncertainties in the ratio are
due to the parametrization of the proton structure function and
dependence on the model. In this figure, we observe that the
entanglement entropy in deeply inelastic scattering has a limit
bound at high inelasticity due to the COM energies. This boundary
value for the entanglement entropy increases with increasing
COM energy of colliders and also shifts to the large value of $Q^2$
region as observed in the EIC, HERA and LHeC colliders in Figs. 4
and 7.\\
\begin{figure}[h]
\includegraphics[width=0.6\textwidth]{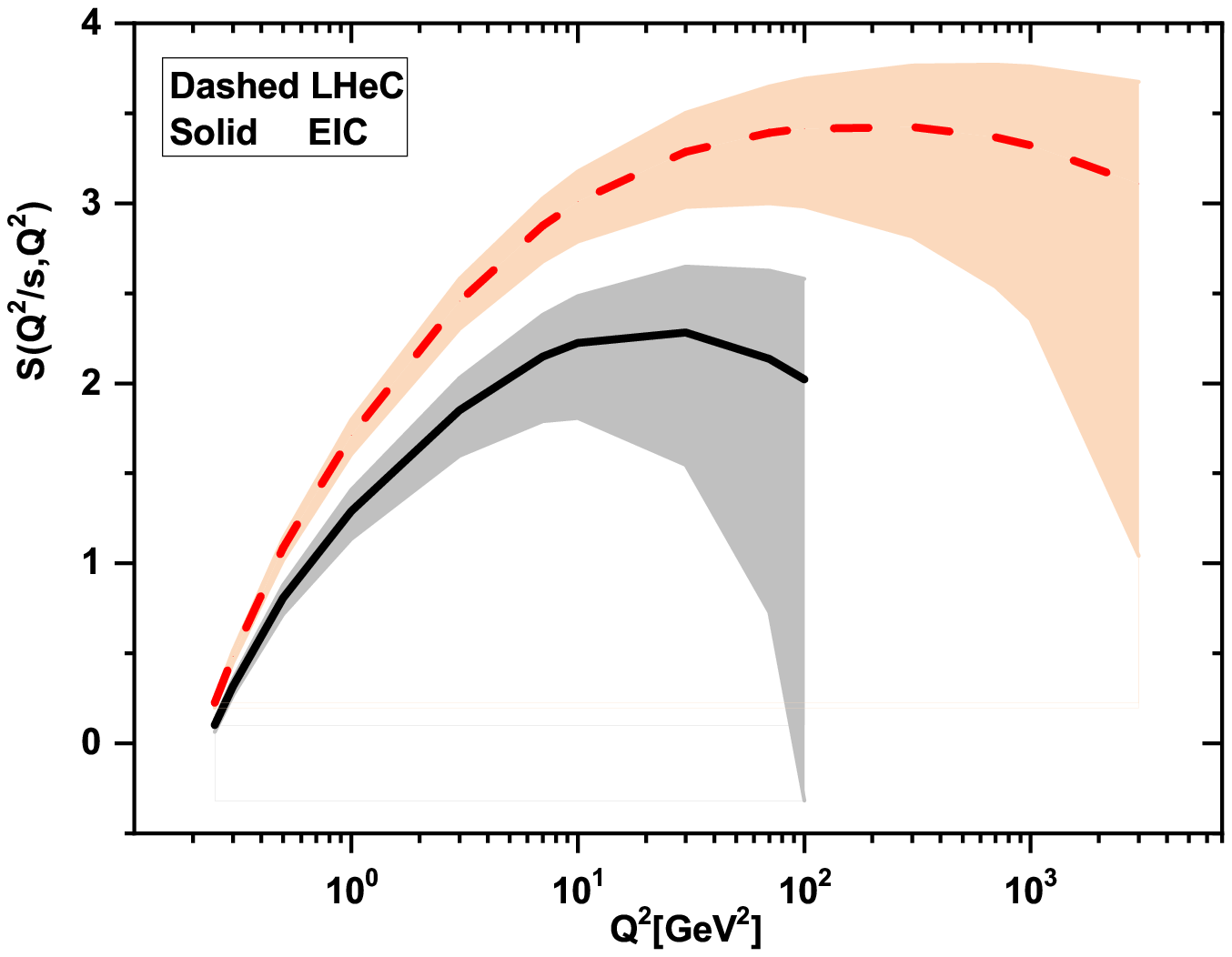}
\caption{ We plotted the entropy $S(\frac{Q^2}{s}, Q^2)$ as a
function of $Q^2$ at $y=1$ for the EIC (black-solid curve) and the
LHeC (red-dashed curve) COM energies with
$\sqrt{s}=89~\mathrm{GeV}$ and $1.3~\mathrm{TeV}$ respectively.
The error
 bands represent the uncertainty in the
parametrization of $F_{2}$ in \cite{Martin1}. }\label{Fig7}
\end{figure}

In conclusion, this study examines the entanglement entropy in a
momentum-space approach to deeply inelastic scattering using the
DIS structure functions, comparing it to existing literature where
entropy has been expressed in terms of parton distribution
functions. Additionally,using physical structure functions
provides greater transparency in  parametrizing initial conditions
for evolution. This method involves parametrizing the function
$F_{2}(x,Q^2)$ within a kinematical region characterized by low
values of the Bjorken variable $x$. Our findings show that the
entanglement entropy for charged hadrons in momentum space
accurately reflects the behavior of the extracted $S(x,Q^2)$, and
our results for the entropy of charged hadrons are comparable to
data from the H1 Collaboration and other results obtained using
the Regge-like
ansatz.\\
This method demonstrates the derivative of entropy behavior in
relation to the Bjorken value $x$ at a fixed $Q^2$, following the
H1 data. It shows good consistency when compared to the HSS and
HERAPDF methods. We then determined the entanglement entropy
for charged hadrons at $y=1$ where
$x_{\mathrm{Bj}}=x_{\mathrm{min}}=Q^2/s$ in the H1 data. The
behavior of the entanglement entropy for charged hadrons shows
that as  $Q^2{\rightarrow}0$, $S(Q^2/s,Q^2)$ also approaches 0. At this
kinematic point, the polarization of the virtual photon is
transverse, and an effective intercept  dominates the derivatives of
$S(x_{\mathrm{min}}, sx_{\mathrm{min}})$ with respect to
$\ln{1/x_{\mathrm{min}}}$ in the high inelasticity limit.\\
The HT corrections to the proton structure function by adding a
simple form $F_{2}{\ast}H_{2}/Q^2$ to  the entanglement entropy
$S(\frac{Q^2}{s}, Q^2)$ at low-$x$ and low-$Q^2$ values are
considered. These effects show that the entanglement entropy
increases as the $Q^2$ value increases at high inelasticity due to
the increasing  COM energy of colliders. The entanglement
entropy $S(\frac{Q^2}{s}, Q^2)$ at the EIC and the LHeC COM
energies are considered in a wide range of $Q^2$.\\

\subsection{Appendix A}
The coefficient functions read as
\begin{eqnarray}
C_{2}&=&\widehat{A}_{2}+\frac{8}{3}a_{s}(Q^{2})DA_{2}\nonumber\\
C_{1}&=&\widehat{A}_{1}+\frac{1}{2}\widehat{A}_{2}+\frac{8}{3}a_{s}(Q^{2})D[A_{1}+(4\zeta_{2}-\frac{7}{2})A_{2}]\nonumber\\
C_{0}&=&\widehat{A}_{0}+\frac{1}{4}\widehat{A}_{2}-\frac{7}{8}\widehat{A}_{2}+\frac{8}{3}a_{s}(Q^{2})D[A_{0}
+(2\zeta_{2}-\frac{7}{4})A_{1}\nonumber\\
&&+(\zeta_{2}-4\zeta_{3}-\frac{17}{8})A_{2}],
\end{eqnarray}
\begin{eqnarray}
\widehat{A}_{2}&=&\widetilde{A}_{2}\nonumber\\
\widehat{A}_{1}&=&\widetilde{A}_{1}+2DA_{2}\frac{\mu^{2}}{\mu^{2}+Q^{2}}\nonumber\\
\widehat{A}_{0}&=&\widetilde{A}_{0}+DA_{1}\frac{\mu^{2}}{\mu^{2}+Q^{2}}\nonumber\\
\widetilde{A}_{i}&=&\widetilde{D}A_{i}+D\overline{A}_{i}\frac{Q^{2}}{Q^{2}+\mu^{2}}\nonumber\\
\widetilde{D}&=&\frac{M^{2}Q^{2}[(2-\lambda)Q^{2}+\lambda
M^{2}]}{[Q^{2}+M^{2}]^{3}}\nonumber\\
\overline{A}_{\varepsilon}&=&a_{\varepsilon1}+2a_{\varepsilon2}L_{2},~~a_{02}=0.
\end{eqnarray}
and
\begin{eqnarray}
\widehat{B}^{(1)}_{L,s}&=&8C_{F}[\frac{25}{9}n_{f}-\frac{449}{72}C_{F}+(2C_{F}-C_{A})\nonumber\\
&&(\zeta_{3}+2\zeta_{2}-\frac{59}{72})]\nonumber\\
\overline{B}^{(1)}_{L,s}&=&\frac{20}{3}C_{F}(3C_{A}-2n_{f})\nonumber\\
\widehat{\delta}^{(1)}_{sg}&=&\frac{26}{3}C_{A}\nonumber\\
\overline{\delta}^{(1)}_{sg}&=&3C_{F}-\frac{347}{18}C_{A}\nonumber\\
\widehat{R}^{(1)}_{L,g}&=&-\frac{4}{3}C_{A}\nonumber\\
\overline{R}^{(1)}_{L,g}&=&-5C_{F}-\frac{4}{9}C_{A}\nonumber\\
L_{A}&=&L+\frac{A_{1}}{2A_{2}}\nonumber\\
L_{C}&=&L+\frac{C_{1}}{2C_{2}}\nonumber\\
L&=&\ln(1/x)+L_{1}\nonumber\\
L_{1}&=&{\ln}\frac{Q^{2}}{Q^{2}+\mu^{2}}\nonumber\\
L_{2}&=&{\ln}\frac{Q^{2}+\mu^{2}}{\mu^{2}}\nonumber\\
A_{i}(Q^{2})&=&\sum_{k=0}^{2}a_{ik}L_{2}^{k},~ (i=1,2)\nonumber\\
A_{0}&=&a_{00}+a_{01}L_{2}\nonumber\\
D&=&\frac{Q^{2}(Q^{2}+\lambda M^{2})}{(Q^{2}+M^{2})^2},
\end{eqnarray}
with the color factors $C_{A}=3$ and $C_{F}=\frac{4}{3}$
associated with the color group $SU(3)$ and $n_{f}$ being the
number of flavors.\\
\begin{table}
\caption{ The effective parameters of $ F_2(x,Q^2)$  at small $x$
for $0.15~\mathrm{GeV}^{2}<Q^{2}<3000~\mathrm{GeV}^{2}$ provided
by the following values. The fixed  parameters are defined by the
Block-Halzen fit to the real photon-proton cross section as
$M^{2}=0.753 \pm 0.068~ \mathrm{GeV}^{2}$, $\mu^2 = 2.82 \pm
0.290~ \mathrm{GeV}^{2}$, and $a_{00}=0.2550\pm 0.016$
\cite{Martin1}.}
\begin{tabular} {cccc}
\toprule \\  \multicolumn{2}{c}{parameters \quad \quad \quad ~~~~~~~~~~~~~~~~value}    \\ &&&\\ \hline \\ &&&\\
  $a_{10} $  &   \quad  $8.205\times 10^{-4}~~  \pm  4.62\times10^{-4} $  \\

  $a_{11} $  &   \quad   $-5.148\times 10^{-2}\pm 8.19\times10^{-3}$  \\

  $a_{12}$   &    \quad  $-4.725\times 10^{-3}\pm 1.01\times10^{-3}$   \\  &&&\\

 $a_{20}$   &   \quad   $2.217\times 10^{-3}\pm 1.42\times10^{-4} $ \\

 $a_{21}$   &   \quad   $1.244\times 10^{-2}\pm 8.56\times10^{-4}$  \\

 $a_{22}$    &    \quad  $5.958\times 10^{-4}\pm 2.32\times10^{-4} $ \\ &&& \\

$a_{00}$& \quad  $2.550\times 10^{-1}~\pm 1.600\times10^{-2}$ & &\\

$a_{01}$& \quad  $1.475\times 10^{-1}~\pm 3.025\times10^{-2}$ & &\\

\hline

\end{tabular}
\end{table}
\begin{table}
\caption{ The effective parameter of $ C$ and the intercept
$\Delta$ were extracted in  \cite{Kutak2}.}
\begin{tabular} {|c|c|c|c|c|}
\hline
 $Q^2/\mathrm{GeV}^2~\in$ & [5,10] & [10,20] & [20,40] & [40,100] \\ [0.5ex]
 \hline\hline
  $\Delta_{\mathrm{HSS}} $    &   \quad  $0.27 \pm 0.04$&   \quad  $0.28 \pm 0.04$&   \quad  $0.29 \pm 0.05$&   \quad  $0.29 \pm 0.04 $  \\
$\Delta_{\mathrm{HERAPDF}} $  &   \quad  $0.28 \pm 0.08$&   \quad  $0.31 \pm 0.07$&   \quad  $0.33 \pm 0.06$&   \quad  $0.36 \pm 0.06 $  \\
$C_{\mathrm{HSS}} $          &   \quad  $2.1 \pm 0.3$&   \quad  $2.7 \pm 0.4$&   \quad  $3.2 \pm 0.5$&   \quad  $3.9 \pm 0.7 $  \\
$C_{\mathrm{HERAPDF}} $      &   \quad  $2.54 \pm 0.68$&   \quad  $2.62 \pm 0.78$&   \quad  $2.65 \pm 0.91$&   \quad  $2.66 \pm 1.02 $  \\
\hline
\end{tabular}
\end{table}
\subsection {ACKNOWLEDGMENTS}

The authors are thankful to the Razi University for financial
support of this project. Additionally, G.R.Boroun would like to
express thanks to Professor Phuoc Ha for his helpful
comments and invaluable support.\\


\end{document}